\documentclass[aps,prb,groupedaddress,a4paper]{revtex4-2}
\usepackage{graphicx,amsmath,braket}

\begin{document}

\title{Extensions to the Su-Schrieffer-Heeger Model: Linear chains and
  their topological properties}

\author{Dyn Paulo C. Dasallas}\email{dcdasallas@up.edu.ph}
\author{Eduardo C. Cuansing}\email{eccuansing@up.edu.ph}
\affiliation{Institute of Physics, University of the Philippines Los
  Ba\~{n}os, Laguna 4031, Philippines}

\date{11 October 2024}

\begin{abstract}
  The Su-Schrieffer-Heeger (SSH) model describes the dynamics of spinless
  fermions in a one-dimensional lattice, with sublattices $A$ and $B$, and
  governed by staggered hopping potentials $v$ and $w$ representing the
  intracell and intercell hopping energies, respectively. In this study,
  we extend the SSH model into three distinct types: a trimer chain, the
  generalized trimer chain, and a hexagonal chain. The trimer chain
  involves three sublattices with intracell and intercell hopping
  potentials $v$ and $w$, respectively. The generalized trimer chain
  incorporates the intracell hopping $v_1$ and $v_2$ and intercell hopping
  $w_1$ and $w_2$ to differentiate the hopping energies between different
  sublattices in the chain. The hexagonal chain is composed of six
  sublattices with intracell hopping potential $v$ and intercell hopping
  potential $w$. We utilize exact diagonalization to determine the bulk
  eigenvalues of the different models. We find that in the trimer and
  generalized trimer chain, the bulk eigenvalues exhibit conducting
  characteristics, independent of the hopping parameter, owing to the
  presence of a flat band situated along the Fermi energy. In the hexagonal
  chain, the bulk eigenvalues display semi-metallic characteristics in the
  region $v<w$ and metallic when $v=0$. Furthermore, we investigate the
  presence of conducting edge states in the finite chains. The trimer and
  hexagonal chains show the presence of topologically protected edge states
  which are manifestations of one-dimensional topological insulators. We also
  established the bulk-boundary correspondence to calculate for the winding
  number that predicts the existence of localized edge states in the
  topological nontrivial phase. 
\end{abstract}

\maketitle

\section{Introduction}
\label{sec:intro}

Recently, topological insulators gained a lot of attention due to their
unique electrical properties as opposed to regular insulators. Topological
insulators (TI) are materials that have an energy band gap between the
valence band and conducting band but have gapless edge states in one and
two dimensions and surface states for 3D TI's \cite{Asboth2016}. The
different phases of matter can be classified according to Landau’s theory
of symmetry breaking \cite{Hasan2010} which states that the different
phases are due to their differences in symmetry. Furthermore, a phase
transition occurs when there is a transition that changes the symmetry of
the system \cite{Wen2004}. However, the classification of phase transitions
in topological insulators is beyond Landau's theory. This means that TI's
can have different phases in zero temperature without breaking symmetry and
the absence of classical phase transitions \cite{Wen2004}. To differentiate
these phases, we describe them by means of topological order \cite{Guo2016}.
These topological orders are generalized properties of zero temperature
states having a finite band gap and does not change unless the system
passes through a quantum phase transition, which is a singularity in the
ground-state energy as a function of the parameters in the Hamiltonian
\cite{Wen2004}, as the temperature is increased. Furthermore, TI's also
gained attention due to their possible applications in spintronics and
quantum computing \cite{Moore2010}.

\begin{figure}
  \includegraphics[width=0.45\columnwidth,clip]{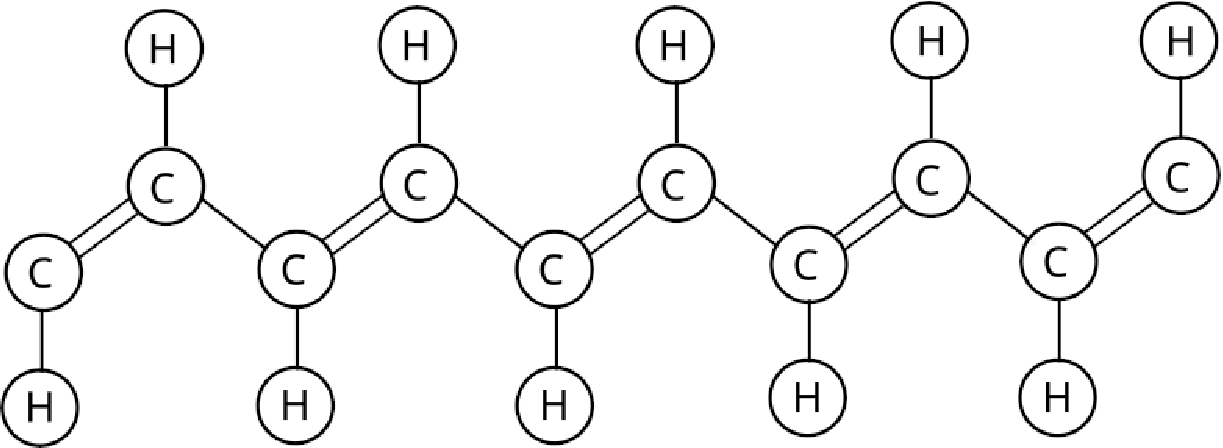}
  \caption{The $trans$ configuration of polyacetylene.
    \label{fig:polyacetylene}}
\end{figure}
  
While most TI's are studied in two and three dimensions, one-dimensional
models give us a good arena to study because of their reduced complexity
and accessibility to experiments \cite{Guo2016}. A good toy model for
understanding topological insulators is the Su-Schrieffer-Heeger (SSH)
model \cite{Su1979}. The SSH model describes the hopping of spinless
fermions in a one-dimensional (1D) lattice with staggered hopping
potentials. It has a topological invariant winding number that shows the
existence of edge states and differentiates the insulating phases through
a quantum phase transition \cite{Asboth2016}. The SSH model was first
applied in the study of the {\it trans} configuration of polyacetylene,
which is the simplest conjugated polymer with alternating single and
double covalent bonds \cite{Su1979}. Each of the carbon atoms in the
chain form four covalent bonds. One is with the hydrogen atom and three
are with the neighboring carbon atoms. The SSH model can be used to find
out the properties of spinless fermions in a chain, such as
trans-polyacetylene, with alternating double and single bonds, represented
by $v$ and $w$, as shown in Fig.~\ref{fig:polyacetylene}.

Recent theoretical extensions of the SSH model include a 1D tripartite
chain \cite{Bercioux2017}, 2D systems in square lattices
\cite{Obana2019} and arm-chair and zigzag graphene nanoribbons
\cite{Fujita1997}. In this study, we extend the 1D SSH model into a
tripartite chain and a hexagonal chain. We then determine the eigenstates
and eigenvalues, i.e., the band structure of the extended models and
establish the bulk-boundary correspondence to find out if topological
edge states are present in the models. We introduce our models in
Sec.~\ref{sec:models} and investigate their corresponding eigenstates,
energy spectra, and winding numbers in Sec.~\ref{sec:eig}. 

\section{Theoretical Models}
\label{sec:models}

\subsection{Trimer Chain}
\label{subsec:trimer}

\begin{figure}[!h]
  \includegraphics[width=0.45\columnwidth,clip]{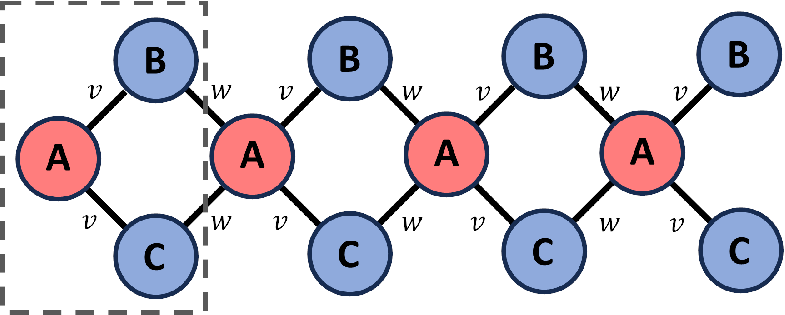}
  \caption{The trimer chain. A single unit cell is composed of 3
    sublattices $A$, $B$, and $C$ and is connected by the hopping energy
    $v$. A unit cell is shown enclosed in a box.
    \label{fig:trimermodel}}
\end{figure}

The trimer chain (see Fig.~\ref{fig:trimermodel}) is composed of three
sublattices $A$, $B$, and $C$ arranged in a diamond chain with staggered
hopping potentials $v$ and $w$ accounting for the intracell and intercell
hopping energies, respectively.The single particle Hamiltonian for this
trimer chain is given by
\begin{widetext}
  \begin{align}
    H = \sum\limits_{m=1}^N v\left(\ket{m,B}\bra{m,A} + \ket{m,C}\bra{m,A}
    + {\rm h.c.}\right)
    + \sum\limits_{m=1}^{N-1} w\left(\ket{m+1,A}\bra{m,B} +
    \ket{m+1,A}\bra{m,C} + {\rm h.c.}\right)
    \label{eq:Htrimer}
  \end{align}
\end{widetext}
where $m$ is the cell index, $v$ and $w$ are the hopping potentials, and
$\rm{h.c.}$ denotes hermitian conjugation. This gives the matrix elements
$H_{ij\alpha\beta} = \braket{i,\alpha|\hat{H}|j,\beta}$, where the indices
$i,j= 1,2,\ldots,N$ denote the cell number and $\alpha,\beta = A,B,C$
denote the sublattice index. We also set the on-site potential to zero,
i.e., $H_{i=j,\alpha=\beta} = 0$. Furthermore, electron-electron
interactions are neglected.

For the bulk, we set the boundary condition to be periodic, i.e., the
Born-von Karman condition. We consider an effectively infinitely long
chain because of the periodic boundary condition. This also implies that
the bulk now is translationally invariant and Bloch’s theorem holds
\cite{Asboth2016}. Our bulk Hamiltonian with the periodic boundary
condition reads,
\begin{widetext}
  \begin{align}
    \begin{split}
      H_{\rm bulk} = & \sum\limits_{m=1}^N v\left(\ket{m,B}\bra{m,A}
      + \ket{m,C}\bra{m,A} + \rm{h.c.}\right)\\
      & + \sum\limits_{m=1}^N w\left(\ket{\left( m\,{\rm mod}\, N\right)+1,A}
      \bra{m,B} + \ket{\left(m\,{\rm mod}\, N\right)+1,A}\bra{m,C}
      + \rm{h.c.}\right)
      \label{eq:Htrimerbulk}
    \end{split}
  \end{align}
\end{widetext}
To account for the periodicity, we introduce terms for the hopping between
sites $1A$ and $NB$, $1A$ and $NC$, and vice versa.

\subsection{Generalized trimer chain}
\label{subsec:generalizedtrimer}

\begin{figure}
  \includegraphics[width=0.45\columnwidth,clip]{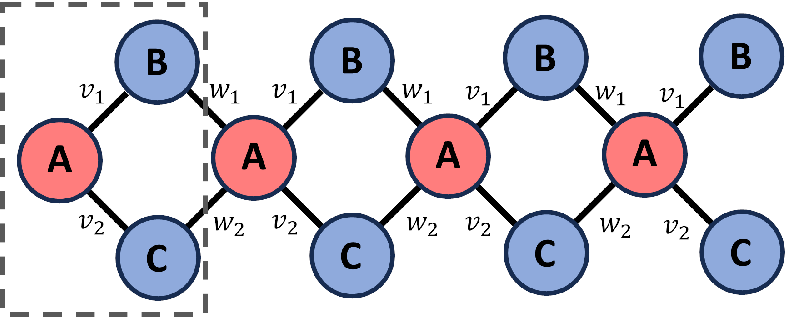}
  \caption{The generalized trimer chain. A single unit cell is composed of
    3 sublattices $A$, $B$, and $C$ and connected by the hopping energies
    $v_1$, $v_2$, $w_1$, and $w_2$. A single unit cell is shown enclosed
    in a box.
    \label{fig:gentrimermodel}}
\end{figure}

The generalized trimer chain is a generalization of the trimer chain
introduced in Sec.~\ref{subsec:trimer}. This model differentiates intracell
and intercell hopping energies for $AC$, i.e., site $A$ to site $C$ in
the same unit cell, and $AB$, i.e., site $A$ to site $B$ in the same unit
cell, and $BA$, i.e., site $B$ to site $A$ in the neighboring unit cell,
and $CA$, i.e., site $C$ to site $A$ in the neighboring unit cell, hopping
energies (see Fig.~\ref{fig:gentrimermodel}). The single-particle
Hamiltonian for this model is
\begin{widetext}
  \begin{align}
    \begin{split}
      H = & \sum\limits_{m=1}^N v_1\left(\ket{m,B}\bra{m,A}
      + {\rm h.c.}\right) + \sum\limits_{m=1}^{N-1}
      v_2\left(\ket{m,C}\bra{m,A} + {\rm h.c.}\right)\\
      & + \sum\limits_{m=1}^{N-1} w_1\left(\ket{m+1,A}\bra{m,B}
      + {\rm h.c}\right) + \sum\limits_{m=1}^{N-1} w_2\left(
      \ket{m+1,A}\bra{m,C} + {\rm h.c}\right)
      \label{eq:Hgentrimer}
    \end{split}
  \end{align}
\end{widetext}
For the bulk part of the chain, we model it as a chain with periodic
boundary conditions. The Hamiltonian for the bulk is 
\begin{widetext}
  \begin{align}
    \begin{split}
      H_{\rm bulk} = & \sum\limits_{m=1}^N v_1 \left(\ket{m,B}\bra{m,A}
      + {\rm h.c.}\right) + \sum\limits_{m=1}^N v_2 \left(\ket{m,C}
      \bra{m,A} + {\rm h.c}\right)\\
      &+ \sum\limits_{m=1}^N w_1 \left(\ket{\left(m\,{\rm mod}\,N\right)+1,A}
      \bra{m,B} + {\rm h.c}\right) + \sum\limits_{m=1}^N w_2\left(
      \ket{\left(m\,{\rm mod}\,N\right)+1,A}\bra{m,C} + {\rm h.c}\right)
      \label{eq:Hgentrimerbulk}
    \end{split}
  \end{align}
\end{widetext}
The additional terms in the Hamiltonian account for the periodicity of the
bulk.

Similar to the work of Bercioux et al. \cite{Bercioux2017}, we calculated
the band structures of a tripartite lattice with infinite and finite unit
cells. Lattice I of their work is similar to our trimer chain with
intercell hopping in the $A$, $B$, and $C$ sublattices. Our generalized
trimer chain generalizes the tripartite system which includes their
Lattice II as a special case when $v_1 = w_2$ and $v_2 = w_1$. To highlight
the novelty of our work, we will show the eigenstates of the finite chains
in the topological nontrivial and trivial regime. We will also calculate
the winding number to establish the bulk-boundary correspondence of our
three model chains.

\subsection{Hexagonal chain}
\label{subsec:hexagonal}

\begin{figure}
  \includegraphics[width=0.72\columnwidth,clip]{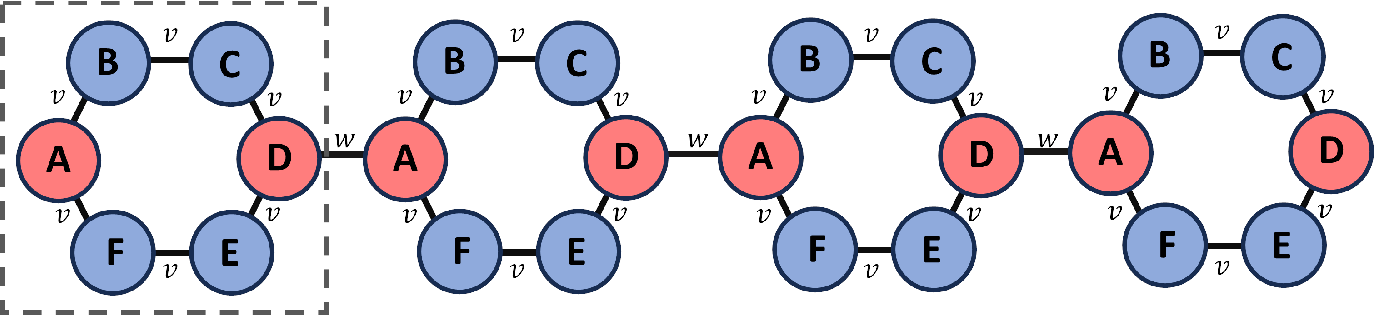}
  \caption{The hexagonal chain. A single unit cell is composed of 6
    sublattices $A$, $B$, $C$, $D$, $E$, and $F$ and is connected by the
    hopping energy $v$ and intercell hopping energy $w$. A unit cell is
    shown enclosed by the box.
    \label{fig:hexamodel}}
\end{figure}

The hexagonal chain forms a resemblance to a 1D graphene armchair ribbon,
as shown in Fig.~\ref{fig:hexamodel}. It consists of six sublattices and
staggered hopping potentials $v$ and $w$ for the intracell and intercell
hopping, respectively. The single-particle Hamiltonian for this hexagonal
chain is
\begin{widetext}
  \begin{align}
    \begin{split}
      H = & \sum\limits_{m=1}^N v \left(\ket{m,B}\bra{m,A}
      + \ket{m,C}\bra{m,B} + \ket{m,D}\bra{m,C} + \ket{m,D}\bra{m,E}
      + \ket{m,E}\bra{m,F}\right.\\
      & + \left.\ket{m,F}\bra{m,A} + {\rm h.c.}\right)
      + \sum\limits_{m=1}^{N-1} w\left(\ket{m+1,A}\bra{m,D}
      + {\rm h.c.}\right)
      \label{eq:Hhexagon}
    \end{split}
  \end{align}
\end{widetext}
where $m$ is the unit cell index. The elements of the Hamiltonian matrix
are given by $H_{ij\alpha\beta} = \braket{i,\alpha|\hat{H}|j,\beta}$ where
$i,j = 1,2,\ldots,N$ and $\alpha,\beta = A,B,C,D,E,F$. The Hamiltonian
for the bulk part of the chain modeled using periodic boundary conditions
is
\begin{widetext}
  \begin{align}
    \begin{split}
      H_{\rm bulk} = & \sum\limits_{m=1}^N v \left(\ket{m,B}\bra{m,A}
      + \ket{m,C}\bra{m,B} + \ket{m,D}\bra{m,C} + \ket{m,D}\bra{m,E}
      + \ket{m,E}\bra{m,F}\right.\\
      & + \left.\ket{m,F}\bra{m,A} + {\rm h.c.}\right)
      + \sum\limits_{m=1}^N w \left(\ket{\left(m\,{\rm mod}\,N\right)+1,A}
      \bra{m,D} + {\rm h.c.}\right)
      \label{eq:Hhexagonbulk}
    \end{split}
  \end{align}
\end{widetext}

\section{Eigenvalues, eigenstates, and the winding number}
\label{sec:eig}

\subsection{Trimer chain}
\label{subsec:eigtrimer}

\begin{figure}[!h]
  \includegraphics[width=0.85\columnwidth,clip]{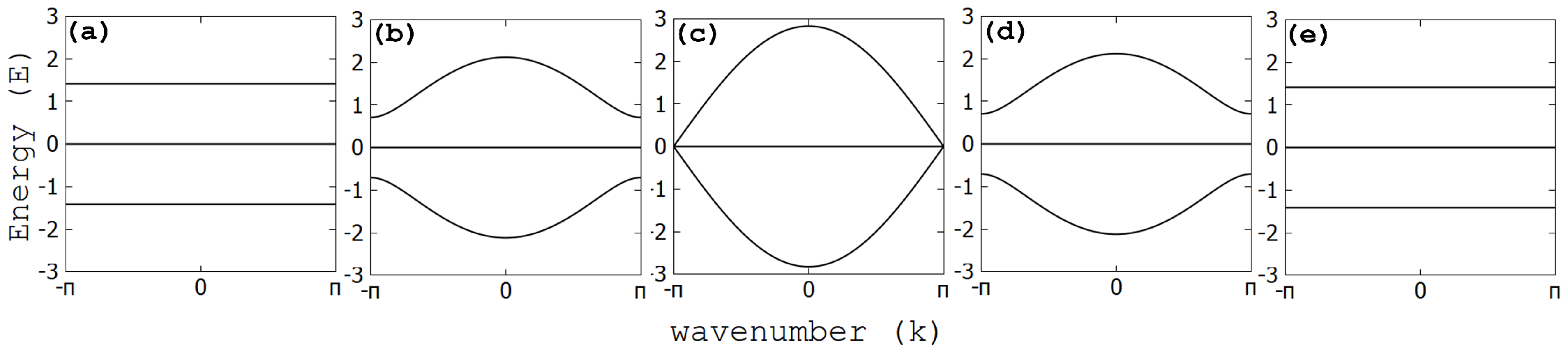}
  \caption{Bulk eigenvalues of the trimer chain plotted in the first
    Brillouin zone ($k = [-\pi,\pi]$) with Fermi energy $E_f = 0$.
    (a) $v=1$, $w = 0$. (b) $v=1$, $w=0.5$. (c) $v=1$, $w=1$.
    (d) $v-0.5$, $w=1$. (e) $v=0$, $w=1$. 
    \label{fig:trimerdispersion}}
\end{figure}

Considering the bulk Hamiltonian of the trimer chain,
Eq.~(\ref{eq:Htrimerbulk}), the energy eigenvalues of the system can be
determined using exact diagonalization. The Schr\"{o}dinger equation for
the bulk Hamiltonian in matrix form is given by
\begin{align}
  \left( \begin{array}{ccc}
    0 & v+w e^{-ik} & v+w e^{-ik}\\
    v+w e^{ik} & 0 & 0\\
    v+w e^{ik} & 0 & 0
  \end{array}\right)
  \left( \begin{array}{c}
    A(k)\\ B(k)\\ C(k)
  \end{array}\right) = E(k)
  \left( \begin{array}{c}
    A(k)\\ B(k)\\ C(k)
  \end{array}\right).
  \label{eq:trimereigenvalueeqn}
\end{align}
This gives the energy eigenvalues
\begin{align}
  E_0(k) = 0~~~{\rm and}~~~E_{\pm}(k) = \pm\sqrt{2\left(v^2+w^2
    + 2vw\cos k\right)}
  \label{eq:trimereigenvalues}
\end{align}
where $k$ is the wave number taking up values in the first Brillouin zone
\cite{Asboth2016}. There are three eigenvalues in
Eq.~(\ref{eq:trimereigenvalues}). The first eigenvalue (zero energy) has
energy coinciding with the Fermi Energy, i.e., $E_0(k) = 0 = E_f$
\cite{Asboth2016}. By varying the magnitude of the hopping parameters $v$
and $w$, a dispersion relation shown in Fig.~\ref{fig:trimerdispersion} can
be obtained.

The dispersion relation consists of five different scenarios concerned
with the different magnitudes of the hopping parameter.
Fig.~\ref{fig:trimerdispersion}(a) describes the system with $v=1$ and
$w=0$. Examining the figure, it can be seen that there exists a flat band
situated at the Fermi level $E=0$. This is due to $E_0(k)$ which has a
constant value and is independent on the wavenumber $k$. This
characteristic can be seen in all the configurations indicating that all
of the configuration of the system is a conductor \cite{Simon2013}.
Furthermore, two flat bands can also be observed at regions
$E = \pm\sqrt{2} v$. The same observations can be seen in
Fig.~\ref{fig:trimerdispersion}(e) with flat bands situated at
$E=\pm\sqrt{2} w$ for $v=0$ and $w=1$.  The case for $v>w$,
Fig.~\ref{fig:trimerdispersion}(b), and $v<w$,
Fig.~\ref{fig:trimerdispersion}(d), show sinusoidal variations of the
energy with respect to the wave number. A defined gap between the
uppermost and the lowest energy band can be seen. Lastly, for the case
where $v=w$, Fig.~\ref{fig:trimerdispersion}(c), the uppermost and the
lowest energy band crossed in the region $k=\pm\pi$ indicating an
intersection, at a point, between the two bands. 

\begin{figure}[!h]
  \includegraphics[width=0.75\columnwidth,clip]{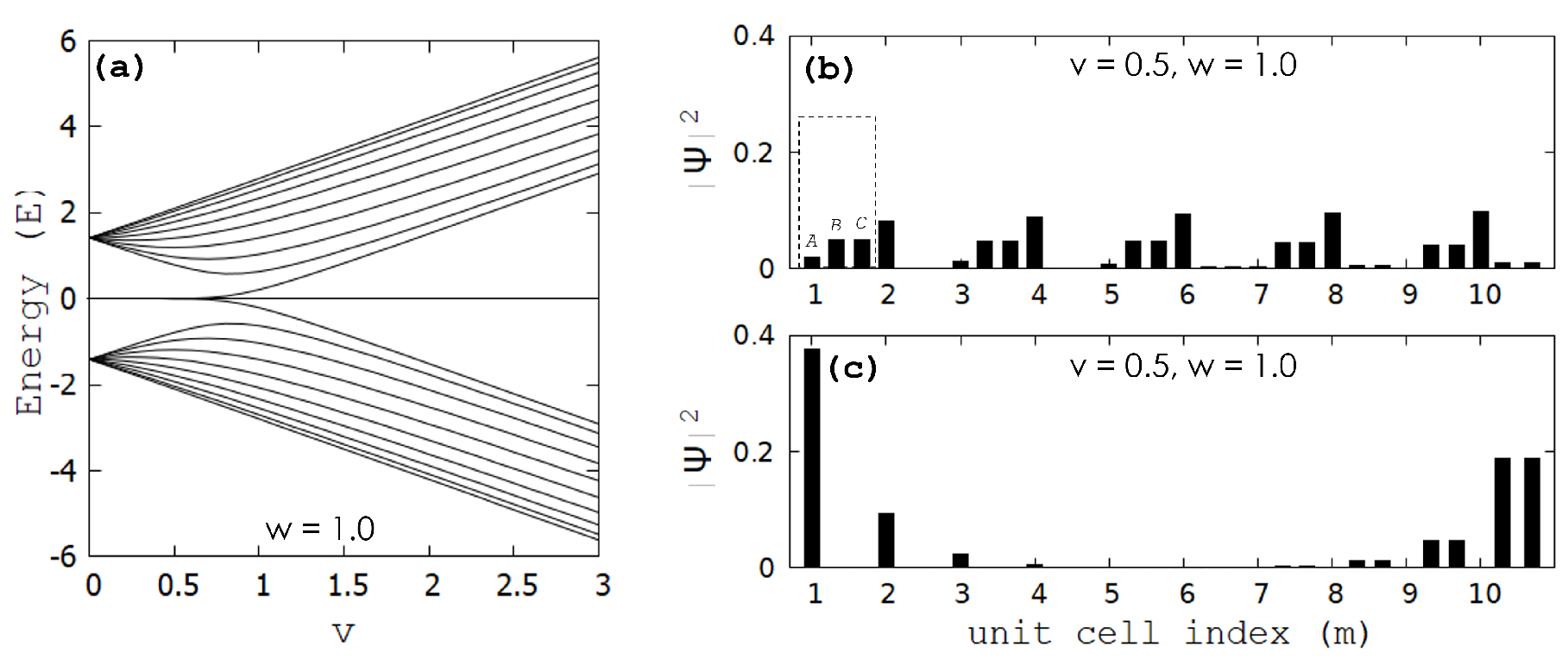}
  \caption{Eigenvalues and eigenstates of a finite trimer chain with
    $N=10$. (a) Energy of $N=10$ unit cell trimer chain varying $v$ while
    keeping $w=1$. (b) Probability density of an arbitrary nonzero energy
    state. A single unit cell is enclosed by the dashed line with
    sublattice $A$, $B$, $C$. (c) Probability density of a zero-energy edge
    state.
    \label{fig:trimerspectra}}
\end{figure}

To consider the characteristics of the boundaries of the chain, we examine
a finite trimer chain. For the extreme cases where $v=0$ or $w=0$  the chain
breaks down into trimers. For the trivial case $w=0$, it can be observed
that all of the sites belong to a corresponding trimer. However, for the
topological case where $v=0$, there are sites that do not belong to a
corresponding trimer. Since fermions in these sites do not hop, then these
edge sites host zero-energy states given by
$\hat{H}\ket{1,A} = \hat{H}\ket{N,B} = \hat{H}\ket{N,C} = 0$.

To calculate for the eigenvalues of the finite trimer chain, we diagonalize
the Hamiltonian given by
\begin{align}
  H = \left(\begin{array}{ccccccc}
    A & B & 0 & \cdots\\
    B^{\rm T} & A & B & 0 & \cdots\\
    0 & B^{\rm T} & A & B & 0 & \cdots\\
    \cdots & & & & & & \cdots\\
    & \cdots & 0 & B^{\rm T} & A & B & 0\\
    & & \cdots & 0 & B^{\rm T} & A & B\\
    & & & \cdots & 0 & B^{\rm T} & A
  \end{array}\right)
  \label{eq:Htrimermatrix}
\end{align}
with
\begin{align}
  A = \left(\begin{array}{ccc}
    0 & v & v\\
    v & 0 & 0\\
    v & 0 & 0
  \end{array}\right)~~~{\rm and}~~~
  B = \left(\begin{array}{ccc}
    0 & 0 & 0\\
    w & 0 & 0\\
    w & 0 & 0
  \end{array}\right),
\end{align}
where for $N=10$ unit cells, the Hamiltonian is a $30\times30$ matrix. By
varying the hopping amplitude $v$ of the finite trimer chain, the
eigenvalue spectra can be obtained (see Fig.~\ref{fig:trimerspectra}). As
observed from Fig.~\ref{fig:trimerspectra}(a), zero-energy states span
throughout the graph independent of the values of $v$. This corresponds to
the zero-energy eigenvalue that we calculated earlier for the bulk (see
Eq.~(\ref{eq:trimereigenvalues})). Notice that these states with zero energy
are degenerate. In Fig.~\ref{fig:trimerspectra}(a), there are $10$ states
having the same zero energy. Furthermore, in the region where $v<w$, there
exists two additional zero-energy states that reside at the edges of the
chain. These results are consistent with those found by Bercioux et al.
\cite{Bercioux2017}. This region is defined as the topologically nontrivial
phase as opposed to the region where $v>w$, which we define as the
topologically trivial phase. As we will see later in this section, these
regions have different topological invariants that distinguishes their
topological phase. Additionally, a topological phase transition can be
observed in the region where the hopping potentials are equal.
Fig.~\ref{fig:trimerspectra}(b) and (c) show the probability density
$|\Psi|^2$ for the corresponding unit cells. Fig.~\ref{fig:trimerspectra}(b)
shows an arbitrary state where $E\neq 0$ wherein the probability density
extends throughout the whole chain. In contrast,
Fig.~\ref{fig:trimerspectra}(c) shows a state where $E=0$. As seen in this
plot, the probability density is localized on both the left and right edges
of the chain. These states are defined as edge states, where the wave
function is localized on both edges of the chain and exponentially
decaying into the bulk. These edge states are typical characteristics of
a one-dimensional topological insulator \cite{Asboth2016}. 

\subsection{Generalized trimer chain}
\label{subsec:eiggeneralized}

\begin{figure}[!h]
  \includegraphics[width=0.85\columnwidth,clip]{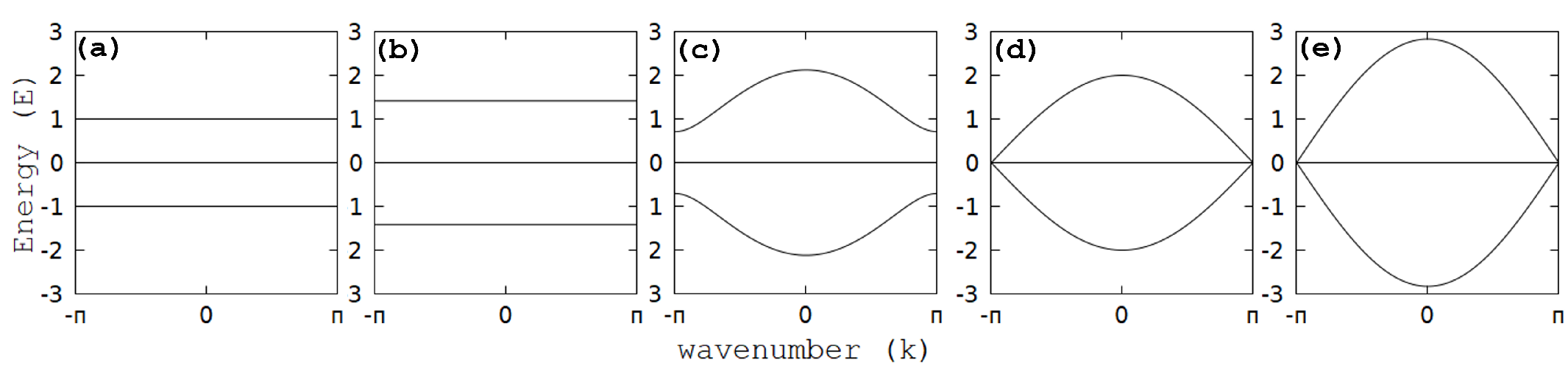}
  \caption{Bulk eigenvalues of the generalized trimer chain plotted in
    the first Brillouin zone ($k=[-\pi,\pi]$) with Fermi energy $E_f=0$.
    (a) $v_1=1$, $v_2=w_1=w_2=0$. (b) $v_1=v_2=0$, $w_1=w_2=1$.
    (c) $v_1=0$, $v_2=w_1=w_2=1$. (d) $v_1=w_1=0$, $v_2=w_2=1$.
    (e) $v_1=v_2=w_1=w_2=1$.
  \label{fig:gentrimerdispersion}}
\end{figure}

Following the same procedure as in the previous subsection, we solve for
the energy eigenvalues of the generalized trimer chain using exact
diagonalization. The Schr\"{o}dinger equation, using the bulk Hamiltonian
in matrix representation, is given by
\begin{align}
  \left( \begin{array}{ccc}
    0 & v_1 + w_1~e^{-ik} & v_2 + w_2~e^{-ik}\\
    v_1 + w_1~e^{ik} & 0 & 0\\
    v_2 + w_2~e^{ik} & 0 & 0
  \end{array}\right)
  \left( \begin{array}{c}
    A(k)\\ B(k)\\ C(k)
  \end{array}\right) = E(k)
  \left(\begin{array}{c}
    A(k)\\ B(k)\\ C(k)
    \end{array}\right).
  \label{eq:gentrimereigeqn}
\end{align}
This gives the energy eigenvalues
\begin{align}
  \begin{split}
    E_0(k) & = 0,\\
    E_{\pm}(k) & = \pm\sqrt{v_1^2 + w_1^2 + 2 v_1 w_1 \cos k + v_2^2 + w_2^2
      + 2 v_2 w_2 \cos k},
    \label{gentrimereigenvalues}
  \end{split}
\end{align}
Notice that when we set $v_1=v_2$ and $w_1=w_2$, we will arrive at the
same energy eigenvalues as those for the trimer chain in
Eq.~(\ref{eq:trimereigenvalues}). Shown in Fig.~\ref{fig:gentrimerdispersion}
are the plots of the energy eigenvalues of the generalized trimer chain
with respect to the wave number $k$. As seen from the plots, different
combinations of the hopping potentials lead to the same conducting
characteristics due to the flat band situated along the Fermi level. This
band is due to the $E_0(k)$ that we obtained. Furthermore, as expected, the
configurations where $v_1=v_2$ and $w_1=w_2$ resemble the dispersion
relations of the trimer chain shown in Fig.~\ref{fig:trimerdispersion}(c).
A visible gap between the uppermost and lowest energy band can be seen in
Fig.~\ref{fig:gentrimerdispersion}(c) in the case where one hopping
parameter is zero ($v_1=0$) while the rest are equal. It can be
generalized that this band gap will remain open until we reach the limit
where $v_1=w_1$ or $v_2=w_2$ or both (see
Fig.~\ref{fig:gentrimerdispersion}(d) and (e)).

\begin{figure}
  \includegraphics[width=0.8\columnwidth,clip]{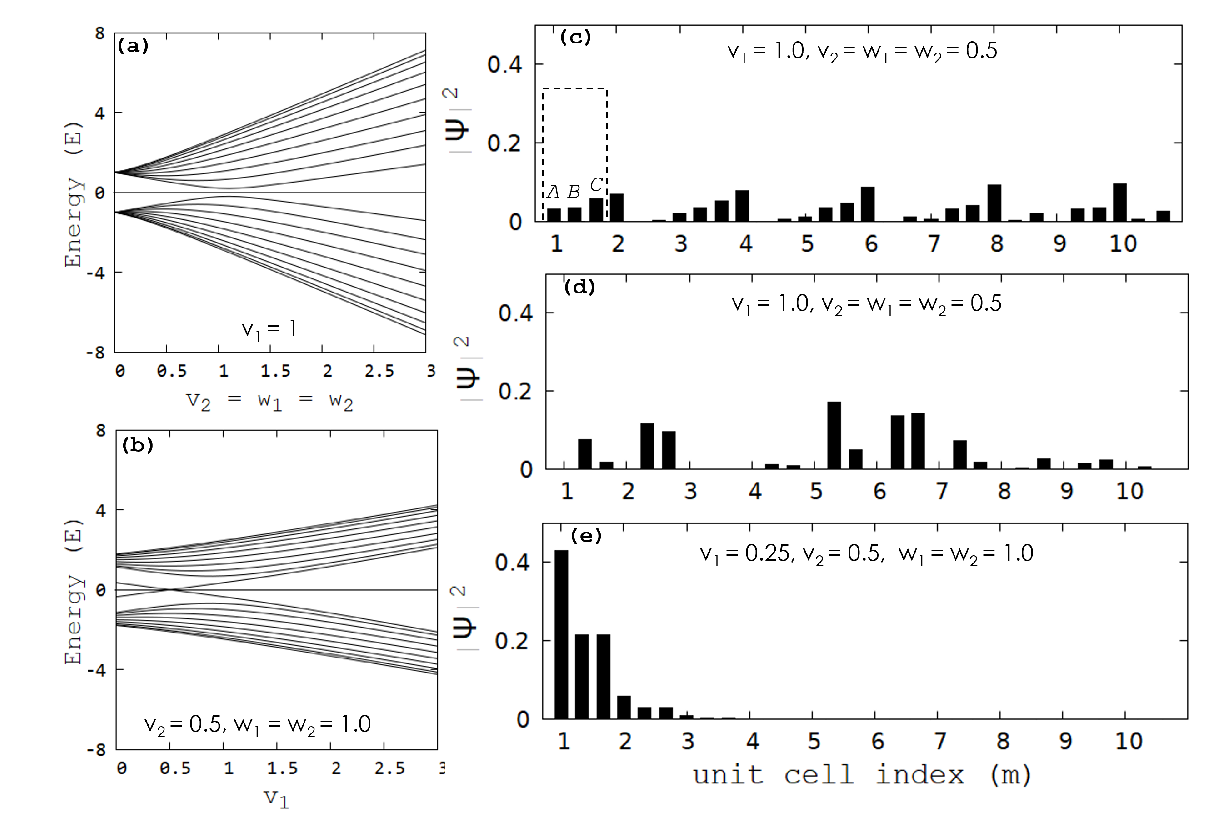}
  \caption{Eigenvalues and eigenstates of a finite generalized trimer chain
    with $N=10$. (a) Energy of $N=10$ unit cells generalized trimer chain
    with varying $v_2=w_1=w_2$ while keeping $v_1=1$. (b) Energy of $N=10$
    unit cells generalized trimer chain with $v_2=0.5$ and $w_1=w_2=1$ and
    with varying $v_1$. (c) Probability density of an arbitrary non-zero
    energy state. A single unit cell is enclosed by a dashed line with
    sublattice $A,B,C$. (d) Probability density of a zero-energy
    degenerate state. (e) Probability density of a chiral edge state.
    \label{fig:gentrimerspectra}}
\end{figure}

To determine the characteristics of the finite generalized trimer chain,
we consider a lattice composed of $10$ unit cells. We find that when only
one of the hopping potentials is nonzero and the rest are set to zero,
the chain breaks down into dimers and the isolated sites would host
zero-energy states. To calculate for the eigenvalues of the generalized
trimer chain, we diagonalize the Hamiltonian given by
\begin{align}
  H = \left(\begin{array}{ccccccc}
    A & B & 0 & \cdots\\
    B^{\rm T} & A & B & 0 & \cdots\\
    0 & B^{\rm T} & A & B & 0 & \cdots\\
    \cdots & & & & & & \cdots\\
    & \cdots & 0 & B^{\rm T} & A & B & 0\\
    & & \cdots & 0 & B^{\rm T} & A & B\\
    & & & \cdots & 0 & B^{\rm T} & A
  \end{array}\right)
  \label{eq:Hgentrimermatrix}
\end{align}
with
\begin{align}
  A = \left(\begin{array}{ccc}
    0 & v_1 & v_2\\
    v_1 & 0 & 0\\
    v_2 & 0 & 0
  \end{array}\right)~~~{\rm and}~~~
  B = \left(\begin{array}{ccc}
    0 & 0 & 0\\
    w_1 & 0 & 0\\
    w_2 & 0 & 0
  \end{array}\right),
\end{align}
where for a $N=10$ unit cell of the generalized  trimer chain, the
Hamiltonian is a $30\times 30$ matrix. Fig.~\ref{fig:gentrimerspectra}(a)
shows the eigenvalue spectra of the finite $N=10$ generalized trimer chain
by varying the three parameters $v_2=w_1=w_2$ simultaneously while
maintaining $v_1=1$. Also, the topological nontrivial phase in the region
$v<w$, which we found earlier in the trimer chain (see
Fig.~\ref{fig:trimerspectra}(a)), motivated us to explore the generalized
case where $\left(v_1\neq v_2\right)<\left(w_1=w_2\right)$. This case is
shown in Fig.~\ref{fig:gentrimerspectra}(b) where we vary the value of
$v_1$ in the range $[0,3]$ while maintaining $v_2=0.5$ and $w_1=w_2=1$. As
noticed from Fig.~\ref{fig:gentrimerspectra}(b), the states which form the
edges in the trimer chain are now gapped in the region where $v_1<v_2$ and
$v_1>v_2$. This gap only closes in the trimer chain limit where
$\left(v_1=v_2\right)<\left(w_1=w_2\right)$.

Fig.~\ref{fig:gentrimerspectra}(c) shows a representative probability
density located at non-zero energy levels. Fig.~\ref{fig:gentrimerspectra}(d)
shows the probability density of one of the degenerate states in the
generalized trimer chain. Lastly, we examined the eigenstates in the region
where $v_1<v_2$, of which a representative probability density is shown in
Fig.~\ref{fig:gentrimerspectra}(e). Although these states are non-zero
energy gapped states, it displays a unique characteristic where the
probability density is only localized in one of the edges. They are known
as chiral edge states \cite{Alvarez2019}, which resemble only half of the
full edge states of the original SSH model. As we will show later in this
section, these chiral edge states do not share the same winding number as
the gapless zero-energy edge states observed in the trimer chain.

To establish the bulk-boundary correspondence of the trimer and generalized
trimer chains, we express their bulk momentum-space Hamiltonian in terms of
the basis states. Consider the following traceless and hermitian matrices,
which are four of the matrices known as Gell-Mann matrices representing the
basis of the group SU(3) \cite{Haber2017},
\begin{align}
  \lambda_1 = \left(\begin{array}{ccc}
    0 & 1 & 0\\
    1 & 0 & 0\\
    0 & 0 & 0\\
  \end{array}\right),~~~
  \lambda_2 = \left(\begin{array}{ccc}
    0 & -i & 0\\
    i & 0 & 0\\
    0 & 0 & 0\\
  \end{array}\right),~~~
  \lambda_4 = \left(\begin{array}{ccc}
    0 & 0 & 1\\
    0 & 0 & 0\\
    1 & 0 & 0\\
  \end{array}\right),~~~
  \lambda_5 = \left(\begin{array}{ccc}
    0 & 0 & -i\\
    0 & 0 & 0\\
    i & 0 & 0\\
  \end{array}\right).
  \label{eq:gellmann}
\end{align}
The bulk momentum-space Hamiltonian of the trimer chain can then be
expressed as,
\begin{align}
  H(k) = \sqrt{2}\left(v + w\cos k\right) \Lambda_x + \sqrt{2}\left(w \sin k
  \right) \Lambda_y
  \label{eq:bmHtrimer}
\end{align}
in which the new basis $\Lambda_x$ and $\Lambda_y$ are linear combinations of
the Gell-Mann matrices,
\begin{align}
  \begin{split}
    \Lambda_x & = \frac{1}{\sqrt{2}}\left(\lambda_1 + \lambda_4\right),\\
    \Lambda_y & = \frac{1}{\sqrt{2}}\left(\lambda_2 + \lambda_5\right),
  \end{split}
  \label{eq:gellmannbasis}
\end{align}
where the normalization coefficients ensure that the basis states follow
the same trace orthonormality condition as that of the Gell-Mann matrices,
${\rm Tr}\left[\Lambda_i \Lambda_j\right] = 2 \delta_{ij}$
\cite{Haber2017}. On the other hand, for the generalized trimer chain, the
bulk momentum-space Hamiltonian can be expressed as,
\begin{align}
  H(k) = \left(v_1 + w_1 \cos k\right) \lambda_1 + \left(v_2 + w_2 \cos
  k\right) \lambda_4 + \left(w_1 \sin k\right) \lambda_2
  + \left(w_2 \sin k\right) \lambda_5.
  \label{eq:bmgenH}
\end{align}
We can see that Eq.~(\ref{eq:bmgenH}) reduces to Eq.~(\ref{eq:bmHtrimer}) in
the limit $v_1=v_2$ and $w_1=w_2$. To reduce the parameters imposed by the
4-dimensional vector of the bulk momentum-space Hamiltonian of the
generalized trimer chain, we set $w_1=w_2=w$, i.e., effectively reducing the
free parameters from four to three. The bulk momentum-space Hamiltonian of
the generalized trimer chain now reads,
\begin{align}
  H(k) = \left(v_1 + w\cos k\right) \lambda_1 + \left(v_2 + w\cos k\right)
  \lambda_4 + \sqrt{2}\left(w\sin k \right) \Lambda_y.
  \label{eq:bmgenHreduced}
\end{align}
Note that the Hamiltonian in Eq.~(\ref{eq:bmgenHreduced}) satisfies chiral
symmetry with the corresponding chiral operator
\begin{align}
  \Gamma = \left(\begin{array}{ccc}
    1 & 0 & 0\\
    0 & -1 & 0\\
    0 & 0 & -1\\
  \end{array}\right),
  \label{eq:chiralop}
\end{align}
where $\Gamma H \Gamma^{\dagger} = -H$ and $\Gamma^2=1$.

To calculate for the topological invariant winding number, we project the
bulk momentum-space Hamiltonians into their corresponding basis states. As
for the Hamiltonian of the trimer chain, Eq.~(\ref{eq:bmHtrimer}), it has
resemblance with that of the bulk momentum-space Hamiltonian of the original
SSH dimer. As such, we expect its trajectory in the $\Lambda_x\,\Lambda_y$
space to look like the trajectory of the SSH Hamiltonian in $dx\,dy$ space.
From these trajectories, we can calculate the winding number $\nu$ by
counting the number of times the trajectory orbits around the origin,
i.e., around $\Lambda_x = \Lambda_y = 0$. We determine the following winding
number $\nu$ for the trimer chain,
\begin{align}
  \nu = \left\{\begin{array}{cl}
  0, & v>w\\
  {\rm undetermined}, & v=w\\
  1, & v<w\\
  \end{array}\right.
  \label{eq:nutrimer}
\end{align}
The winding number is a useful tool in predicting the existence of gapless
localized edge states (states where the wave function of the system is
localized on both edges of the chain \cite{Asboth2016}).  As shown in
Eq.~(\ref{eq:nutrimer}), for the case where $v>w$, we found that $\nu=0$,
thus indicating a trivial topology of the system characterized by the
absence of localized edge states. On the other hand, the case where $v<w$
indicates a non-trivial topology where $\nu=1$ indicating the presence of
a pair of gapless localized edge states. Furthermore, the case where $v=w$
leads to an undetermined $\nu$ indicating that the trajectory of $H$ is in
direct contact with the origin. This case entails a topological phase
transition, which shows that the only way to change the winding number of
the trimer chain from $\nu=0$ to $\nu=1$, or vice versa, is by closing the
gap between the highest and lowest band of the trimer chain. 

\begin{figure}
  \includegraphics[width=0.77\columnwidth,clip]{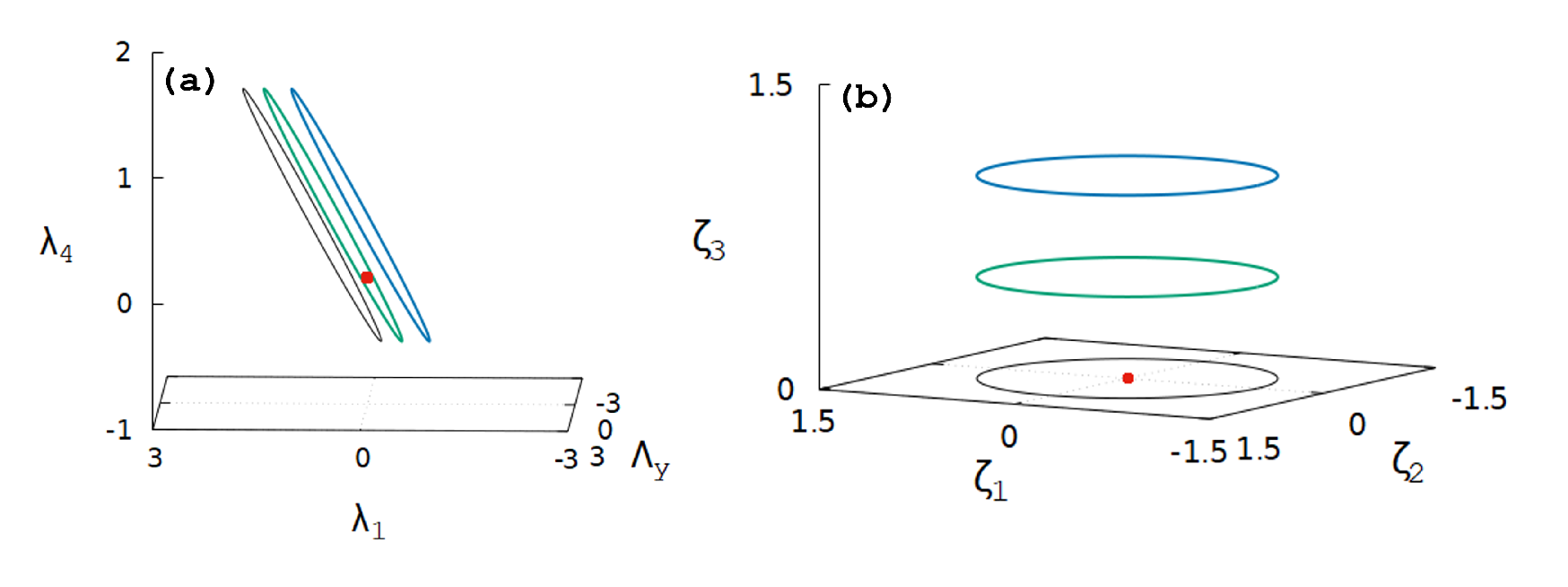}
  \caption{Trajectories of the bulk momentum-space Hamiltonian of the
    generalized trimer chain and hexagonal chain in the $\lambda$ and
    $\zeta$ basis. (a) Trajectories of the bulk momentum-space Hamiltonian
    of the generalized trimer chain with varying $\nu$. The red dot
    corresponds to the origin, i.e., $\lambda_1=\lambda_4=\Lambda_y =0$.
    The blue circle (rightmost) corresponds to the parameters $v_1=0.1$,
    $v_2=0.5$, $w_1=w_2=1$ where the winding number $\nu=0$,the green
    circle (middle) with parameters $v_1=v_2=0.5$, $w_1=w_2=1$ where the
    winding number $\nu=1$ and the black circle (leftmost) with parameters
    $v_1=0.8$, $v_2=0.5$, $w_1=w_2=1$ where the winding number $\nu=0$.
    (b) Trajectories of the bulk momentum-space Hamiltonian of the
    hexagonal chain with varying $\nu$. The red dot corresponds to the
    origin, $\zeta_1=\zeta_2=\zeta_3=0$. The black circle (lowermost)
    corresponds to the parameters $v=0$, $w=1$ where the winding number
    $\nu=1$, the green circle (middle) with parameters $v=0.5$, $w=1$ where
    the winding number $\nu=0$ and the blue circle (uppermost) with
    $v=w=1$ where the winding number $\nu=0$.
    \label{fig:windingnumber}}
\end{figure}

Extending our analysis to the generalized case, we plot the trajectory of
the bulk-momentum space Hamiltonian of the generalized trimer chain,
Eq.~(\ref{eq:bmgenHreduced}) in $\lambda$-space, as shown in
Fig.~\ref{fig:windingnumber}(a). Here we are interested to see if the
winding number changes or remains the same for different variations in the
upper, $v_1$, and lower, $v_2$, intracell hopping parameters. We are
particularly interested in this case due to the non-trivial topological
nature of the trimer chain in the $v<w$ regime, as we have shown earlier
in this section. As such, we would like to generalize the conditions for the
existence of gapless localized edge states in the $\left(v_1\neq v_2\right)<w$
regime in terms of their winding numbers. As observed from
Fig.~\ref{fig:windingnumber}(a) we determined the following winding numbers,
\begin{align}
  \nu = \left\{\begin{array}{cc}
  0, & v_1>v_2\\
  1, & v_1=v_2\\
  0, & v_1<v_2\\
  \end{array}\right.
  \label{eq:nugen}
\end{align}
where $w=1$ for all cases. From this result, we can see that only in the
trimer chain limit, i.e., $v_1=v_2$, is the winding number $\nu=1$ where
we have shown a localized edge state in Fig.~\ref{fig:trimerspectra}(c).
It is also the region where the gap closes at $E=0$ between the pair of
positive and negative energy band in Fig.~\ref{fig:gentrimerspectra}(b).
The other configurations, on the other hand, in Eq.~(\ref{eq:nugen}),
with $v_1>v_2$ and $v_1<v_2$, demonstrate a chiral edge state emerging
from a non-zero energy state, which is quite different from the usual
gapless localized edge states where the probability densities are localized
on both edges of the chain \cite{Alvarez2019}. 

\subsection{Hexagonal chain}
\label{subsec:eighexagonal}

\begin{figure}[!h]
  \includegraphics[width=0.85\columnwidth,clip]{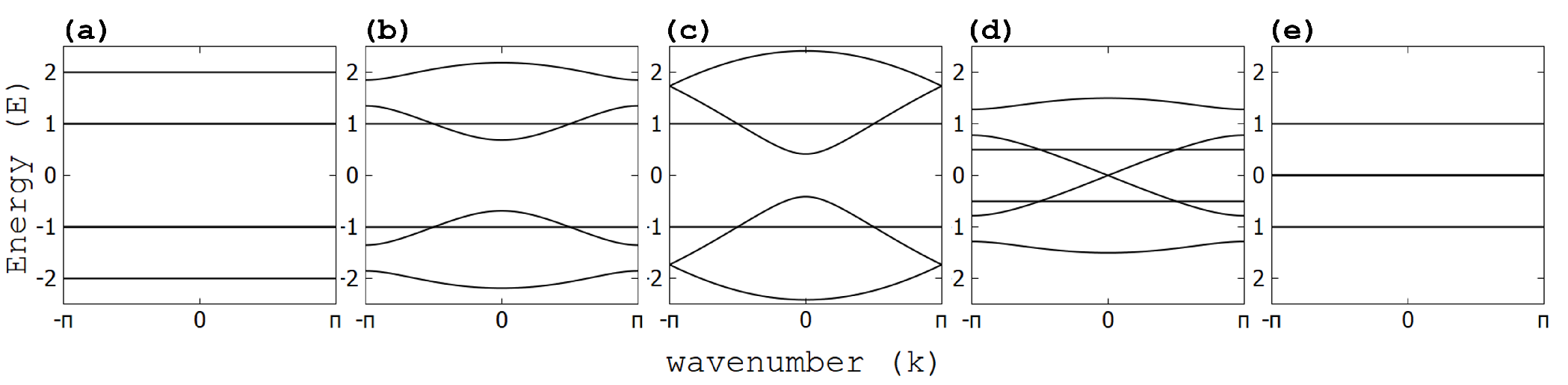}
  \caption{Bulk eigenvalues of the hexagonal chain plotted in the first
    Brillouin zone ($k=[-\pi,\pi]$) with the Fermi energy $E_f=0$.
    (a) $v=1$, $w=0$, (b) $v=1$, $w=0.5$, (c) $v=w=1$, (d) $v=0.5$, $w=1$,
    (e) $v=0$, $w=1$.
    \label{fig:hexeigenvals}}
\end{figure}

Solving for the bulk energy eigenvalues of the hexagonal chain requires us
to solve Schr\"{o}dinger's equation given by
\begin{align}
  \left(\begin{array}{cccccc}
    0 & v & 0 & w e^{-ik} & 0 & v\\
    v & 0 & v & 0 & 0 & 0\\
    0 & v & 0 & v & 0 & 0\\
    w e^{ik} & 0 & v & 0 & v & 0\\
    0 & 0 & 0 & v & 0 & v\\
    v & 0 & 0 & 0 & v & 0\\
  \end{array}\right)
  \left(\begin{array}{c}
    a(k)\\
    b(k)\\
    c(k)\\
    d(k)\\
    e(k)\\
    f(k)\\
  \end{array}\right) = E(k)
  \left(\begin{array}{c}
    a(k)\\
    b(k)\\
    c(k)\\
    d(k)\\
    e(k)\\
    f(k)\\
  \end{array}\right).
  \label{eq:hexschrod}
\end{align}
The energy eigenvalues of the hexagonal chain with varying hopping
parameters can be calculated by numerically diagonalizing the matrix in
Eq.~(\ref{eq:hexschrod}). We do this using the GNU Octave computational
tool software. The eigenvalues are shown in Fig.~\ref{fig:hexeigenvals}.

We see that Fig.~\ref{fig:hexeigenvals}(a) to \ref{fig:hexeigenvals}(c)
show an insulating system due to the presence of a band gap. Furthermore,
in Fig.~\ref{fig:hexeigenvals}(d) where $v<w$, a conical band structure
can be seen at $k=0$. This band structure displays gapless and linear
conducting bands which are typical for a semi-metal \cite{Wan2011}.
Additionally, when $v=0$, see Fig.~\ref{fig:hexeigenvals}(e), the chain
shows conducting characteristics due to the flat band situated along the
Fermi energy, $E_f=0$, independent of the wave number $k$. The eigenvalue
spectra can be determined by diagonalizing the Hamiltonian given by
\begin{align}
  H = \left(\begin{array}{ccccccc}
    A & B & 0 & \cdots\\
    B^{\rm T} & A & B & 0 & \cdots\\
    0 & B^{\rm T} & A & B & 0 & \cdots\\
    \cdots & & & & & & \cdots\\
    & \cdots & 0 & B^{\rm T} & A & B & 0\\
    & & \cdots & 0 & B^{\rm T} & A & B\\
    & & & \cdots & 0 & B^{\rm T} & A
  \end{array}\right)
  \label{eq:Hhexmatrix}
\end{align}
with
\begin{align}
  A = \left(\begin{array}{cccccc}
    0 & v & 0 & 0 & 0 & v\\
    v & 0 & v & 0 & 0 & 0\\
    0 & v & 0 & v & 0 & 0\\
    0 & 0 & v & 0 & v & 0\\
    0 & 0 & 0 & v & 0 & v\\
    v & 0 & 0 & 0 & v & 0\\
  \end{array}\right)~~~{\rm and}~~~
  B = \left(\begin{array}{cccccc}
    0 & 0 & 0 & 0 & 0 & 0\\
    0 & 0 & 0 & 0 & 0 & 0\\
    0 & 0 & 0 & 0 & 0 & 0\\
    w & 0 & 0 & 0 & 0 & 0\\
    0 & 0 & 0 & 0 & 0 & 0\\
    0 & 0 & 0 & 0 & 0 & 0\\
  \end{array}\right),
\end{align}
where for a $N=10$ unit cells hexagonal chain, the Hamiltonian $H$ is a
$60\times 60$ matrix.

\begin{figure}[!h]
  \includegraphics[width=0.75\columnwidth,clip]{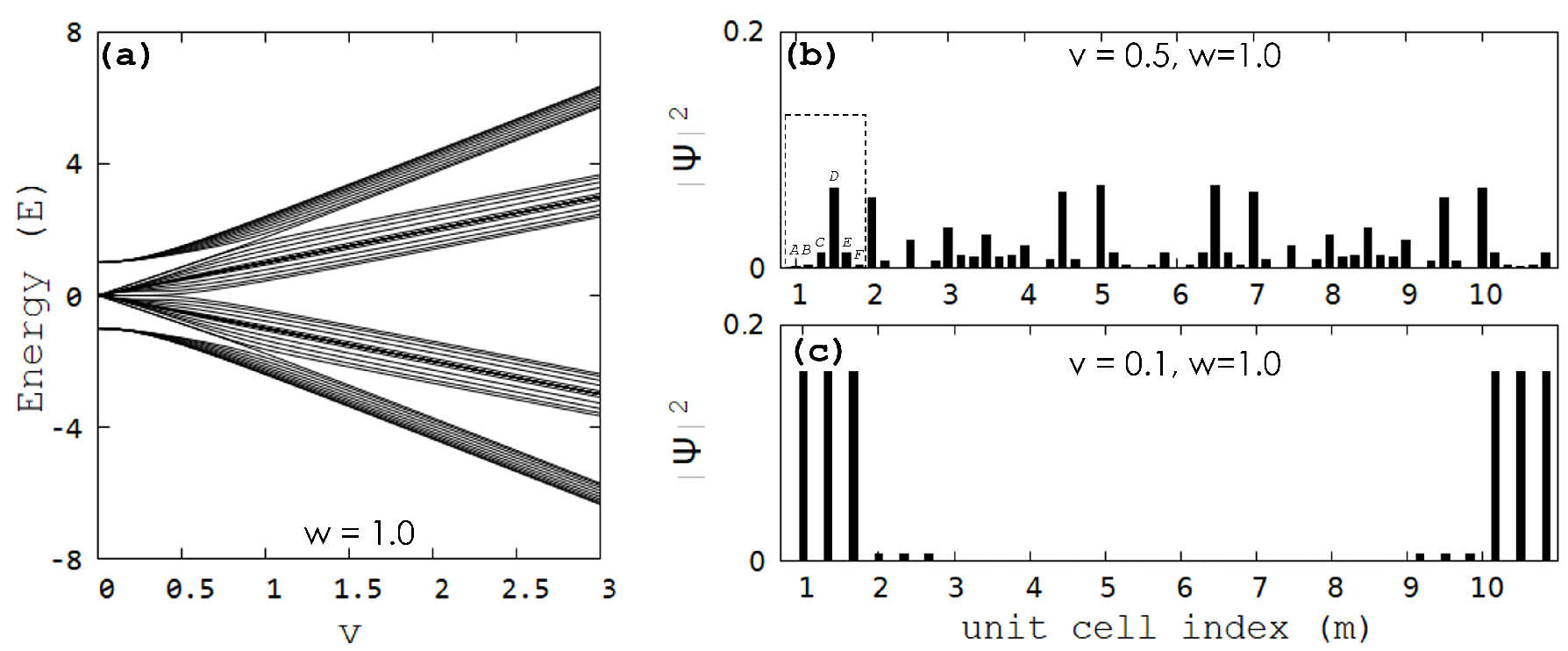}
  \caption{Eigenvalues and eigenstates of a finite hexagonal chain with
    $N=10$. (a) Energy of $N=10$ unit cells hexagonal chain with $w=1$
    and varying $v$. (b) Probability density of a representative nonzero
    energy state. (c) Probability density of a zero-energy edge state.
    \label{fig:hexspectra}}
\end{figure}

Shown in Fig.~\ref{fig:hexspectra}(a) is the eigenvalue spectra obtained
by plotting the energy eigenvalues with $w=1$ and slowly tuning $v$ in
the range $[0,3]$.  Furthermore, Fig.~\ref{fig:hexspectra}(b) and
\ref{fig:hexspectra}(c) show a representative extended state and an
edge state of the finite hexagonal chain. A pair of zero energy edge
states (one is shown in Fig.~\ref{fig:hexspectra}(c)) show up when $v$ is
varied. Additionally, we observe a topological phase transition around
$v=0$. The representative extended state seen in Fig.~\ref{fig:hexspectra}(b)
shows an arbitrary non-zero energy state. In this figure, it can be seen
that the probability density is extended across the chain indicating
the conducting characteristics of a trivial topology. In contrast, the edge
state shown in Fig.~\ref{fig:hexspectra}(c) indicates that the probability
density is localized at the edges of the chain revealing the nontrivial
topology of the system.

To establish the bulk-boundary correspondence of the hexagonal chain, we
express the bulk momentum-space Hamiltonian in terms of its basis states:
\begin{align}
  H(k) = \left(w \cos k\right) \zeta_1 + \left(w \sin k\right) \zeta_2
  + \left(\sqrt{6}v\right) \zeta_3,
  \label{eq:bmHhexa}
\end{align}
where $\zeta_1$, $\zeta_2$, and $\zeta_3$ are
\begin{align}
  \zeta_1 = \left(\begin{array}{cccccc}
    0 & 0 & 0 & 1 & 0 & 0\\
    0 & 0 & 0 & 0 & 0 & 0\\
    0 & 0 & 0 & 0 & 0 & 0\\
    1 & 0 & 0 & 0 & 0 & 0\\
    0 & 0 & 0 & 0 & 0 & 0\\
    0 & 0 & 0 & 0 & 0 & 0\\
  \end{array}\right),~~~
  \zeta_2 = \left(\begin{array}{cccccc}
    0 & 0 & 0 & -i & 0 & 0\\
    0 & 0 & 0 & 0 & 0 & 0\\
    0 & 0 & 0 & 0 & 0 & 0\\
    i & 0 & 0 & 0 & 0 & 0\\
    0 & 0 & 0 & 0 & 0 & 0\\
    0 & 0 & 0 & 0 & 0 & 0\\
  \end{array}\right),~~~
  \zeta_3 = \frac{1}{\sqrt{6}} \left(\begin{array}{cccccc}
    0 & 1 & 0 & 0 & 0 & 1\\
    1 & 0 & 1 & 0 & 0 & 0\\
    0 & 1 & 0 & 1 & 0 & 0\\
    0 & 0 & 1 & 0 & 1 & 0\\
    0 & 0 & 0 & 1 & 0 & 1\\
    1 & 0 & 0 & 0 & 1 & 0\\
    \end{array}\right).
  \label{eq:zetas}
\end{align}
These matrices satisfy the orthonormality condition
${\rm Tr}\left[\zeta_i \zeta_j\right] = 2 \delta_{ij}$. They are 3 of the 35
matrices representing the basis of the group SU(6). Furthermore, we define
the chiral symmetric operator,
\begin{align}
  \Gamma = \left(\begin{array}{cccccc}
    1 & 0 & 0 & 0 & 0 & 0\\
    0 & -1 & 0 & 0 & 0 & 0\\
    0 & 0 & 1 & 0 & 0 & 0\\
    0 & 0 & 0 & -1 & 0 & 0\\
    0 & 0 & 0 & 0 & 1 & 0\\
    0 & 0 & 0 & 0 & 0 & -1\\
  \end{array}\right),
  \label{eq:chihexa}
\end{align}
where $\Gamma H \Gamma^{\dagger} = -H$ and $\Gamma^2=1$.  This $6\times 6$
chiral operator ensures that the energies of the system form chiral
symmetric pairs.

The trajectories of the bulk momentum-space Hamiltonian of the hexagonal
chain in $\zeta$-space, Eq.~(\ref{eq:bmHhexa}) with varying hopping
parameters are shown in Fig.~\ref{fig:windingnumber}(b). The trajectory of
Eq.~(\ref{eq:bmHhexa}) forms a cylindrical plot with radius $w$ and height
$v$. Thus, increasing the value of $w$ and $v$ will increase the radius
and height, respectively. The same as the procedure introduced in the
previous section, we determine the winding number to be
\begin{align}
  \nu = \left\{\begin{array}{cl}
  1, & v=0\\
  0, & v=0.5\\
  0, & v=1\\
  \end{array}\right.,
  \label{eq:nuhexa}
\end{align}
where we keep the value of $w=1$. As seen from the calculation of the
winding number, we expect a pair of localized edge states in the region
$v=0$ where the winding number $\nu=1$. An example of this edge state is
shown in Fig.~\ref{fig:hexspectra}(c).

\section{Conclusion}
\label{sec:conclusion}

In summary, we extended the Su-Schrieffer-Heeger model into three distinct
types: the trimer chain, the generalized trimer chain, and the hexagonal
chain. The trimer chain is a one-dimensional lattice with sites $A$, $B$,
and $C$ and staggered hopping potentials $v$ and $w$ for the intracell
and intercell hopping energies. The generalized trimer chain is a
generalization of the trimer chain with the hopping parameters $v_1$, $v_2$,
$w_1$, and $w_2$ to distinguish the $A$ to $B$ from the $A$ to $C$ hopping.
Lastly, the hexagonal chain which is composed of six sublattices and
staggered hopping potentials $v$ and $w$ for the intracell and intercell
hopping.

Exact diagonalization was used to calculate the energy eigenvalues of the
bulk and the finite systems of the extended models. For the trimer chain,
we find that the bulk characteristic of the chain exhibits conducting
characteristics due to the flat band situated along the fermi energy
independent of the values of $v$ and $w$. Furthermore, when the hopping
potentials are equal, it is seen that the uppermost and lowermost bands
meet at $k=\pm\pi$. The trimer chain's finite structure shows the presence
of gapless zero-energy edge states as well as a topological phase
transition, properties that resemble the characteristic of a one-dimensional
topological insulator. The bulk-boundary correspondence for the trimer chain
also predicts the existence of these localized edge states through
the winding number $\nu$ where we determine $\nu=1$ in the topological
($v<w$) regime.  The bulk characteristic of the generalized trimer chain,
on the other hand, also shows conducting characteristic due to the flat
band situated along the Fermi level. In the generalized case where
$v_1\neq w_1$ and/or $v_2\neq w_2$, a visible band gap can be seen between
the uppermost and the lowest bands. This band gap only closes at the limit
when $v_1=w_1$ and $v_2=w_2$. The bulk-boundary correspondence of the
generalized trimer chain reveals that the gapless topological edge states
(localized on both edges) are only present in the trimer chain limit when
$v_1=v_2$. Gapped chiral edge states on the other hand, i.e., localized
only on the left or right boundary, show up in the region where
$\left(v_1\neq v_2\right)<\left(w_1=w_2\right)$. Lastly, for the hexagonal
chain, the dispersion relations show that the model is a semi-metal when
$v<w$ due to the absence of a band gap and a metal when $v=0$ due to the
presence of degenerate flat bands at the Fermi level. The existence of
gapless localized edge states is also observed in the finite hexagonal
chain at around $v=0$ where the system is in the topological phase defined
by a winding number $\nu=1$.

\begin{acknowledgments}
The authors are grateful to Rafael Bautista for insightful discussions.
\end{acknowledgments}

\bibliography{references}

\end{document}